# Reversible hydrogen storage by controlled buckling of graphene layers

*Valentina Tozzini\* and Vittorio Pellegrini*

NEST Istituto Nanoscienze-CNR and Scuola Normale Superiore,

Piazza San Silvestro 12, 56127 Pisa (Italy)

*To whom the correspondence should be addressed. Email tozzini@nest.sns.it, Phone +39 050 509 433 Fax +39 050 509 417

ABSTRACT

We propose a multi-layer graphene-based device in which storage and release of atomic hydrogen is obtained by exploiting and controlling the corrugation of the layers. The proposal is based on density-functional simulations of hydrogen chemisorption on graphene with superimposed corrugation. We report a tunability of the binding energies of more than 2 eV by changing the sheet out-of-plane deformation up to ± 0.2 Å, the convex regions allocating the energetically favored hydrogen binding sites. We discuss an optimized process of curvature inversion that can lead to efficient and fast hydrogen release and $H_2$ formation. Our corrugated graphene device can potentially reach gravimetric capacities of ~8 wt % and reversibly store and release atomic hydrogen by application of external fields without the need of modifying temperature and pressure.

MANUSCRIPT TEXT

Hydrogen-based fuel cells are promising solutions for the efficient production of electricity for mobile applications. A key step for the development of a reliable technology based on hydrogen fuel-cells requires to solve the issue of storage of molecular hydrogen. Hydrogen-storage systems must be light, inexpensive, robust and display a high hydrogen-storage capacity with a gravimetric density of around 6wt% or more[1]. In addition the hydrogen storage medium must display fast sorption/desorption kinetics and good reversibility.

A number of proposals based on the design of advanced materials such metal hydrides and carbon-based structures have been made to overcome the limitations of the conventional solution of compressing hydrogen in tanks[2,3,4]. Recently, several studies focussed on graphene, the one-atom-thick membrane of carbon atoms packed in a honeycomb lattice, demonstrating on theoretical grounds the possibility to achieve the desired gravimetric capacity of 8-9 wt % sets by the DoE for 2015 for both molecular and atomic hydrogen by designing complex three dimensional patterns or by decorating graphene by appropriate atoms[5,6].

Hydrogen chemisorption on graphene and graphite surfaces has been extensively studied both experimentally[7,8,9,10,11,12] and theoretically[13,14,15,16,17,18,19,20,21,22,23]. It has been shown that chemisorption of atomic hydrogen on flat graphene is an exothermic process with a barrier of $\approx 0.3$ eV and a binding energy of $\approx 0.8$ eV depending on coverage[24]. The range of these values sets limits to the applicability of mechanisms of gas release based on heating or on lowering the hydrogen-gas pressure. However it was also shown that convex curved graphene structures such as nanotubes and fullerenes display an increase of H binding energy leading to vanishing barrier energy to H chemisorption. These calculations suggest novel routes to tune the chemisorption process on curved graphene surfaces and control the amount and pattern of chemisorbed hydrogen[25,26]. This kind of tunability might provide a platform to develop a new technology for hydrogen storage devices.

Here we propose a scheme for a device to store and release hydrogen that exploits buckling of graphene layers. This novel mechanism is made possible by a specific relationship between the hydrogen binding energy and the curvature of the graphene sheet obtained from a density functional theory (DFT) analysis. Our analysis is performed on a supercell containing 180 carbon atoms in a multilayer configuration (shown in Figure 1) in which the interlayer separation is set at ~15 Å to allow efficient hydrogen flow between the layers. In practice, such a separation could be induced by properly designed molecular intercalants.

We find a linear relationship between the binding energy and the curvature level of the carbon atom site, with the binding energy increasing of about 2.25 eV passing from concave to convex regions. Based on these results we define the working scheme of the hydrogen storage device in which hydrogen release is linked to the curvature inversion of the graphene layers. We anticipate that a mechanism for controlled inversion of the graphene layers can be obtained by using appropriate polar or charged chemical intercalates sensitive to external electric/magnetic fields.

These studies indicate a novel route towards the realization of optimized graphene-based hydrogen storage devices in which the C-H interaction can be finely controlled by external fields that act on the mechanical properties of graphene rather then by changes in temperature or pressure.

Figure 1(a) shows the graphene multilayer structure considered here. Our simulations are carried out in the approximately square supercell containing 180 C atoms highlighted in Fig.1(a). The system is periodic also in the z direction, with a 15 Å interlayer separation. This ensures minimal interaction energy between sheets and allows efficient inter-plane hydrogen flow.

Following previously established schemes for C-H systems[27], the calculations were performed within the DFT frame, with Kohn-Sham orbitals expanded in plane waves, with Troullier-Martins pseudopotentials and Perdew-Burke-Ernzerhof energy functional[28]. The molecular dynamics calculations were performed within the Car-Parrinello approach[29] using the CPMD3.13 code[30]. Structure relaxations were obtained with simulated annealing procedures and standard local minimization approaches. The rippled structures were obtained using supercells contracted of 10% of their relaxed size in the directions x, y or both (denominated Rx, Ry and Rxy, respectively). The starting rippled structure was obtained superimposing a sinusoidal displacement from planarity (in x, y or xy) having the supercell periodicity, followed by a structure relaxation with constrained cell size. While the structures Rx and Ry maintain the starting symmetry, the Rxy structure undergoes a global rearrangement during the relaxation phase. The final structure follows the graphene symmetry, with ripples and valleys following the zig-zag lines of the hexagonal geometry. The corrugated structures are reported in Figure 1(b) and (d).

Our simulations show that the corrugation is stable. This is in agreement with recent calculations showing that a lateral compression of 10% is compatible with the spontaneous formation of ripples whose period is of the order of the size cell we have chosen[25]. In the simulation the constraints are periodic by construction, but as it will be clear in the following, the periodicity or regularity of the corrugation is not a requirement for the proper functioning of the hydrogen storage device, provided the curvature level and sign (concave or convex) can be controlled.

Overall the C atoms in the rippled structures span a wide range of different levels of local curvature, as shown by the shades of color. There are several ways to measure the local curvature, or local distortion from planarity. This property is defined for each single C atom by its location and those of its three ligands (first neighbors). One possibility is to consider the improper dihedral angle $\phi$ formed by the vertex and the three basal C atoms, which assumes positive or negative values depending on convexity/concavity. Since the complex formed by the four C atoms is not always regular, a convenient measure is the average value of the improper dihedrals: $\phi = 1/3 \sum_{i=1,3} \phi_i$. The distance d of the vertex C atom from the plane formed by the other three atoms, considered with a sign depending on the side of the plane where it is located, is an alternative measure of the local curvature. The relationship between $\phi$ and d can be analytically evaluated through the formula $d = l \cos(\theta/2) \sin(\phi)$, where $\theta$ is the angle defined by the vertex C and two of its three neighbors and l is the C-C bondlength. The definitions of $\phi$, d and $\theta$ are shown in Figure 1 (c) . The curvature level in structures Rx, Ry and Rxy span a range of d = $\pm 0.25$Å, or equivalently $\phi = \pm 20$deg.

The H binding energy was evaluated relaxing the structure with single H atoms adsorbed on a sample of 35 C sites in locations of different positive and negative local curvature. Figure 2 reports the change of the binding energy $\Delta E_{bind}$ as a function of d and C–H bond-length. $\Delta E_{bind}$ represents the variation with respect to the binding energy of the flat layer ($\Delta E_{bind} = 0$). Our results show that C–H, curvature and $\Delta E_{bind}$ have a good linear correlation, the C–H bond being shorter and stronger in convex regions. In particular the correlation between binding energy and curvature is fairly well expressed by the relationship $\Delta E_{bind}[eV] = -4.499 d[\text{Å}]$ with an error of only 0.02 eV. It can be seen that the binding energy increases of about 2.25 eV passing from concave to convex regions, as an effect of the local distortion of the sp2 hybridization in favor of the sp3. In particular the values for convex graphene regions are in agreement with values found in the literature for fullerenes and nanotubes (red points in Fig.2; the values of d for small nanotubes range around 0.1-0.15 Å while that for fullerene $C_{60}$ is $\approx 0.3$ Å; the maximum value (diamond) is $\approx 0.5$ Å). Thus, we can conclude that by varying graphene curvature the stability of adsorbed H can be tuned by more than 2 eV yielding stable and unstable H adsorbates.

On the basis of the results displayed in Fig.2, we are now in the position to discuss a storage device that exploits the relationship between hydrogen binding energy and graphene mechanical properties. We envision a device constituted by multiple layers of graphene separated by a relatively large distance such that hydrogen can freely flow between the layers. The working scheme of the device requires the layers to be corrugated and their convexity/concavity and lateral com- pression controlled in a correlated way. To this end, the graphene sheet could be appropriately functionalized by means of suitable intercalates, for example bifunctional molecules chemically grafted to two parallel sheets or non-covalent aromatic substituents that experience strong pi-pi interactions with the graphene rings. We note that selective chemical methods for graphene grafting are available for functional groups of different nature (see for example[31,32]). Additionally, graphite was shown to be efficiently exfoliated to graphene sheets by ionic liquids[33], which are a versatile class of molecular systems that can be converted to a number of functional molecular architectures[34]. To account for their function, the intercalates need to respond reversibly to external stimuli such as electro-optic or magnetic by exerting a strain on the graphene sheets to modify their planarity. A promising configuration might exploit symmetric photochromic molecules (e.g. azo-dyes) capable to modify their head-to-tail distance upon light stimulation. Another possible configuration envisions the use of magnetic nanostructures that can be manipulated through an external magnetic field.

A simulation obtained including 180 H atoms in the Rxy system midway between the layers shows that about 50% of the H atoms attach on the convex parts during system relaxation (see Fig.3). We observe that H atoms attach on both sides of the sheet, always on the convex parts. This coverage corresponds to half of the maximum possible gravimetric density of 4%. However it is likely that optimizing the corrugation level this value could be increased up to about 6-8%, corresponding to about 70% of coverage or more. Larger values are not likely because H does not show to spontaneously attach on the flat parts of the structure. The calculated binding energy of this system is about 2.85 eV /H atom, i.e. about 1.5-2eV larger than that of the single isolated adsorbed H atom, and about 0.8-1.5 eV larger than that of the adsorbed "dimers". This large increase of the binding energy is due to three distinct effects (i) the already reported clustering effect, which favors the adsorption of H near already present H atoms; (ii) the fact that H is attached over a convex surface; (iii) the fact that there is a further increase of the convexity due to H binding, already observed for the single H chemiosorbed, that cooperatively increases when many nearby H atoms are chemiosorbed. Thus, overall, the chemiosorption on the convex surfaces is a highly favorable process.

Once H is selectively attached on the convex part we need to efficiently invert the curvature for example exploiting molecular intercalates as described above. In order not to break the graphene sheet this should be done within a proper time dynamics and temporarily releasing the lateral compression of the multilayer, that should be reset immediately afterwords in order to favor the opposite curvature corrugation. Once the convexity of the multilayer is completely inverted, the H atoms found themselves bound to the concave part. This implies a decrease of the binding energy of at least 1.2 eV per atom, which might be increased due to cooperative effects to 2eV. The system therefore becomes unstable at least with respect to the desorbed molecular hydrogen, if not towards the desorbed atomic hydrogen. Selected snapshots of the process are reported in Fig.3. A complete analysis will be reported elsewhere.

In conclusion we have proposed an approach to store and release hydrogen by controlled buck- ling of graphene layers. The working scheme of the device exploits the large sensitivity of the hydrogen binding energy on the curvature of the graphene sheet. With this approach one can envision to build a hydrogen storage device with high gravimetric capacity, stability and fast kinetics that operates at constant temperature and pressure.

ACKNOWLEDGMENT.

We acknowledge the allocation of CINECA supercomputing center resources from INFM Progetto di calcolo Parallelo 2009 and from iit (Italian Institute of Tecnology) - transversal computing platform, 2010. We thank Ranieri Bizzari, Marco Polini and Mikhail Katsnelson for useful discussions and suggestions.

FIGURE CAPTIONS

Figure 1: Description of the model system used in the calculations. (a) Representation of the multilayer system in the flat configuration. The distance between layers is 15 Å. The unit cell (containing 180 C atoms and sized 22.13x21.30 Å) is highlighted. (b) Corrugation of the system is obtained by superimposing a sinusoidal distortion from planarity in x, y or both directions and contracting the cell size accordingly of 10%; (c) Schematic representation of the three parameters (out of plane distance d, vertex angle θ and out of plane dihedral ϕ) that can measure the local curvature. (d) Pictures of the three stable structures for corrugation in dir x (Rx), xy (Rxy) and y (Ry), from top down. Rippled structures are colored according to the curvature (red=convex, blue=concave).

Figure 2: Variation of the binding energy (with respect to the binding energy of the flat geometry) as a function of the C-H bond-length and of the curvature, measured by d. Black circles: our calculations. Red points: available data for nanotubes and fullerenes.

Figure 3: Working scheme of the hydrogen storage device. Three phases can be distinguished: during the injection (red) the atomic hydrogen is introduced in the device, that subsequently chemisorbes on the convex parts initiating the storage phase (blue). The release phase (green) is activated by inversion of curvature, causing the associative desorption of H2. The orizontal lines are placed according to the effective energies/H atom (shades represent error bars). Representa- tive snapshots are reported: the graphene sheet is represented in light brown and the hydrogen in orange.

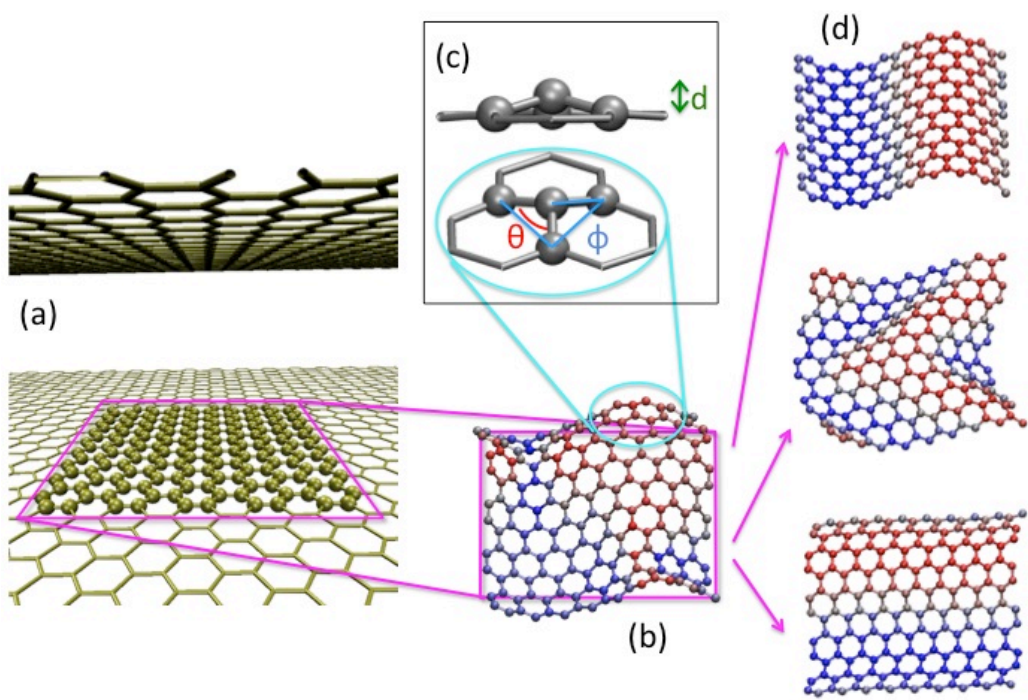

Figure 1

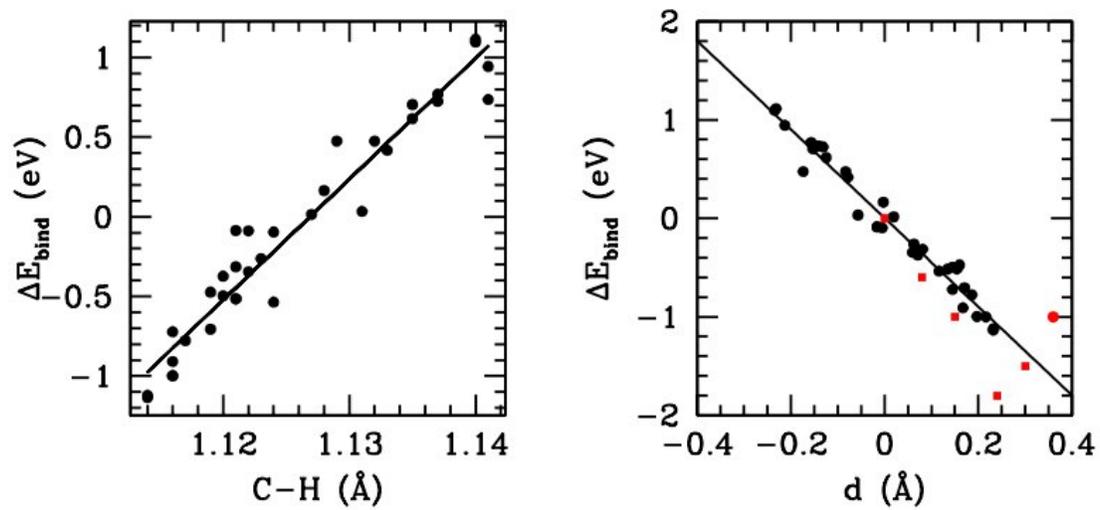

Figure 2

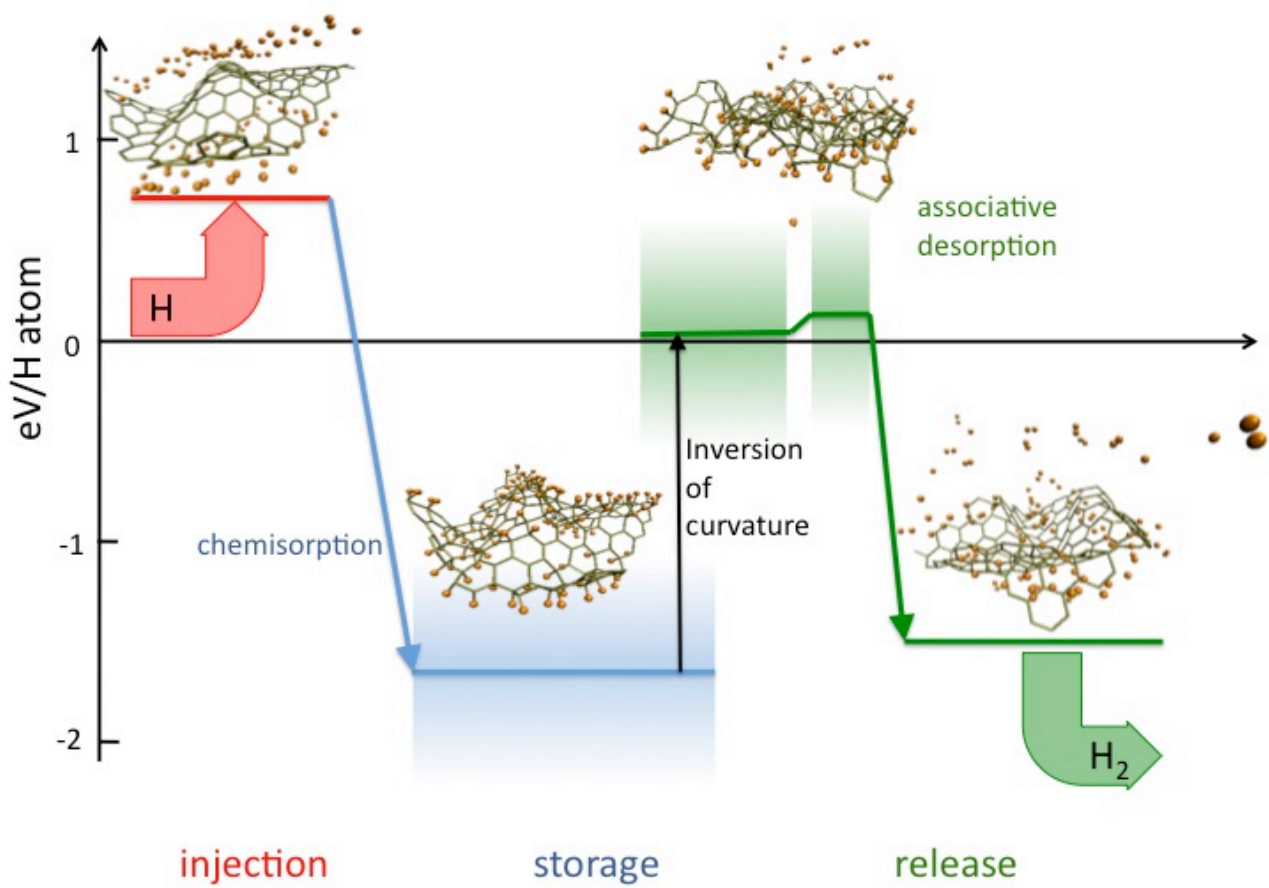

Figure 3

SYNOPSIS TOC

Corrugated graphene structure colored according to the H binding energy (light grey = high binding energy, dark grey= low binding energy, for attachment of H from above)

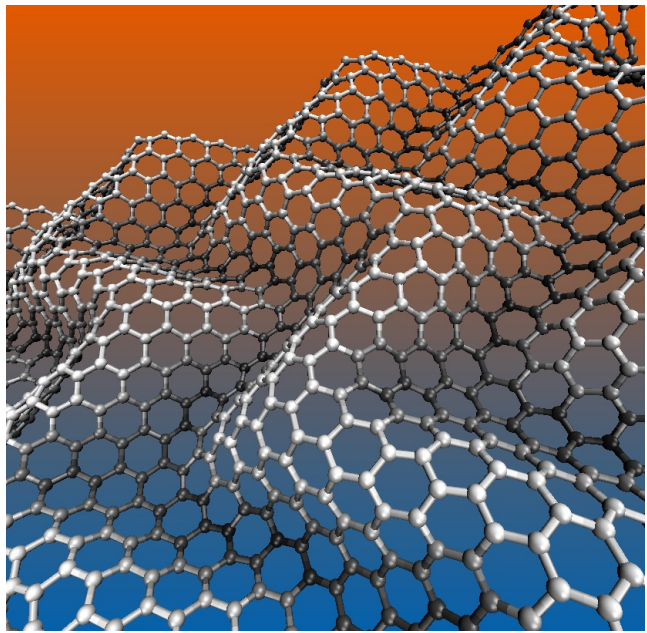

[28] Several tests were performed to optimize the calculation parameters. The cutoff for plane waves was set at 35 ryd but values up to 60 ryd were used to check convergence. The use of BLYP functional in place of PBE did not bring relevant differences in the optimized geometries and energies. Given the large size of the super-cell, the calculations were performed using the Gamma point only. However the geometry optimization and some related calculations were also carried out in a smaller supercell (24 atoms, 7.42x8.58x15 Å) using 5x4x2 K points symmetrically distributed according to Monkhorst-Pack algorithm.